\documentclass[apj]{emulateapj}

\usepackage{amsmath}
\usepackage{float}
\usepackage{booktabs}
\usepackage{hyperref}
\usepackage{verbatim}

\usepackage[caption=false]{subfig}

\usepackage{color}

\newcommand{\pb}{$P_{\rm b}$}
\newcommand{\ma}{$m_{1}$}
\newcommand{\mb}{$m_{2}$}
\newcommand{\ps}{$P_{\rm s}$}
\newcommand{\ecc}{$e$}
\newcommand{\om}{$\omega_{\rm p}$}
\newcommand{\m}{$m$}
\newcommand{\tobs}{$t_{\rm obs}$}
\newcommand{\inc}{$i$}
\newcommand{\gam}{$\gamma_1$}
\newcommand{\gama}{$\gamma_2$}
\newcommand{\gamj}{$\gamma_3$}
\newcommand{\psrpoppy}{{\sc psrpoppy}}

\begin{document}

\title{On the detectability of ultra-compact binary pulsar systems}

\author{Nihan Pol}
\affil{Department of Physics and Astronomy, West Virginia University, Morgantown, WV 26506-6315}
\affil{Center for Gravitational Waves and Cosmology, West Virginia University, Chestnut Ridge Research Building, Morgantown, West Virginia 26505}
\affil{Department of Physics and Astronomy, Vanderbilt University, 2301 Vanderbilt Place, Nashville, TN 37235, USA}
\author{Maura McLaughlin}
\author{Duncan R. Lorimer}
\author{Nathan Garver-Daniels}
\affil{Department of Physics and Astronomy, West Virginia University, Morgantown, WV 26506-6315}
\affil{Center for Gravitational Waves and Cosmology, West Virginia University, Chestnut Ridge Research Building, Morgantown, West Virginia 26505}

\begin{abstract}
    Using neural networks, we integrate the ability to account for Doppler smearing due to a pulsar's orbital motion with the pulsar population synthesis package \psrpoppy\ to develop accurate modeling of the observed binary pulsar population. 
    As a first application, we show that binary neutron star systems where the two components have highly unequal mass are, on average, easier to detect than systems which are symmetric in mass. 
    We then investigate the population of ultra-compact ($1.5 \, {\rm min} \leq P_{\rm b} \leq 15\,\rm min$) neutron star--white dwarf (NS--WD) and double neutron star (DNS) systems which are promising sources for the Laser Interferometer Space Antenna gravitational-wave detector. 
    Given the non-detection of these systems in radio surveys thus far, we estimate a 95\% confidence upper limit of $\sim$1450 and $\sim$1100 ultra-compact NS--WD and DNS systems in the Milky Way that are beaming towards the Earth respectively. 
    %This does not imply fewer ultra-compact DNS systems than NS--WD systems in the Galaxy, but merely that we can place better constraints on the size of the population of the former type of system. 
    We also show that using survey integration times in the range 20~s to 200~s with time-domain resampling will maximize the signal-to-noise ratio as well as the probability of detection of these ultra-compact binary systems.
    Among all the large scale radio pulsar surveys, those that are currently being carried out at the Arecibo radio telescope have $\sim$50--80\% chance of detecting at least one of these systems using current integration integration times and $\sim$80--95\% using optimal integration times in the next several years.
    
    %A detection of just a single ultra-compact binary neutron star system will provide extensive opportunities for multi-messenger science.
\end{abstract}


\section{Introduction}
    
    The era of multi-messenger astronomy was ushered in with GW170817, a detection of gravitational waves (GWs) emitted by the merger of two neutron stars using the Laser Interferometer Gravitational-wave Observatory \citep[LIGO,][]{LIGO_detector_ref}  and Virgo \citep[][]{VIRGO_detector_ref} detectors \citep{GW170817} as well as across the electromagnetic spectrum by a range of ground and space-based telescopes \citep{GW170817_EM}. While LIGO-Virgo has made another confirmed detection of a double neutron star (DNS) merger event \citep{gw190425} and released alerts for a few more potential DNS mergers, these are relatively rare cataclysmic events.
    On the other hand, the Laser Interferometer Space Antenna \citep[LISA,][]{LISA} is a space-based GW detector which is sensitive to compact objects in binary systems   emitting GWs at frequencies between $0.1 \, {\rm mHz} \lesssim f_{\rm GW} \lesssim 100 \, {\rm mHz}$. Given the abundance of binaries consisting of compact objects as well as their non-cataclysmic nature, these systems provide rich potential for long-term multi-messenger science. 
    
    The strongest sources for LISA are Galactic ultra-compact binary (UCB) systems, which are binary systems with stellar-mass components and orbital periods $P_{\rm b} < 15$~min. These UCB systems can consist of any combination of white dwarf, neutron star or black holes, with the most common source ($\sim 10^7$ in the Galaxy) being double white dwarf (DWD) binaries \citep{dwd_pop_synth_2, dwd_pop_synth_1}. However, population synthesis simulations have shown that LISA should also detect a few tens of ultra-compact double neutron star (DNS) and neutron star--white dwarf (NS--WD) systems \citep{lisa_dns_andrews, lau_detecting_dns_w_lisa}. UCB systems are ``verification binaries'' for LISA, i.e. they  should be detectable within weeks of LISA beginning operations. Verification binaries for DWD systems have already been identified in the electromagnetic (EM) band using optical surveys \citep{lisa_dwd_1, lisa_dwd_2, lisa_dwd_3}. However, no verification binary consisting of a neutron star has been detected yet. The most promising DNS system for detection by LISA is the Double Pulsar system, J0737--3039 \citep{0737A_disc}, but this system will only accumulate a signal-to-noise ratio (S/N) of $\sim$3 over the planned 4-yr LISA mission.
    
    Joint, multi-messenger observations of these UCB systems can provide significantly more information than observations in the EM or GW bands alone. As shown by \citet{shah_lisa_mm_inc}, measuring the inclination of an UCB system through EM observations can improve the constraint on the GW amplitude of that system by a factor as large as six. In addition, knowing the sky position of an UCB system can improve the GW parameter estimation by a factor of two \citep{shah_lisa_mm_skypos}. Additionally, for DNS systems, joint EM and GW observations can constrain the mass-radius relation to within $\approx$0.2\% \citep{thrane_lisa_dns}. Thus, it is important to find as many UCB systems as possible before LISA is launched in the 2030s to maximize the scientific potential of the mission.
    
In the EM band, neutron star binaries are discovered by searching for pulsars, which are rapidly rotating neutron stars emitting beamed emission at radio wavelengths. So far, 185 pulsars have been discovered in binary systems with a white dwarf companion, while 20 pulsars have been discovered in binary systems with another neutron star \citep[for an up-to-date list, see the ATNF pulsar catalog\footnote{https://www.atnf.csiro.au/research/pulsar/psrcat/},][]{psrcat}. The shortest orbital period for a pulsar--WD binary is $\sim$2~hours \citep[PSR~J1518+0204C,][]{shortest_pb_wd_1, shortest_pb_wd_2}, while for DNS systems the shortest orbital period is $\sim$1.8~hours \citep[J1946+2052,][]{1946_disc}. Henceforth in this paper, we assume that the binary system contains a pulsar whenever we refer to ultra-compact NS--WD or DNS systems.
    
    The limiting factor in detecting UCB systems in radio-wavelength surveys is the Doppler smearing of the pulsar emission due to its orbital motion \citep{og_odf} as it causes a reduction in the S/N with which the pulsar is detected. This Doppler smearing is quantified using the orbital degradation factor \citep{og_odf}, which can take values between 0 and 1, and lower values of the orbital degradation factor signify higher Doppler smearing and thus a larger reduction in S/N for the pulsar. The orbital degradation factor depends on, among other things, the orbital period of the binary system and is smaller for systems with small orbital periods. Thus, UCB systems, with their extremely small orbital periods, are difficult to detect in normal radio pulsar surveys

    To improve sensitivity to pulsars in binary systems, where the apparent pulse period can change significantly during the observation due to the Doppler effect, acceleration and jerk search techniques are employed in radio pulsar surveys \citep[see][for a review of implementation techniques]{lorimer_kramer}. Acceleration searches have now been widely implemented in the search pipelines for almost all large radio pulsar surveys \citep[for example,][]{accel_search_implementation_1}, while jerk searches are only recently being implemented \citep{jerk_search_implementation} due to the technique being significantly more computationally expensive than acceleration searches. The effect of the acceleration search technique on the S/N of the pulsar was quantified in \citet{og_odf} for circular binaries, while \citet{Bagchi_odf} expanded their work to include eccentric systems as well as the effect of jerk search techniques.
    
    While these techniques have been well known in the literature, they have never been fully incorporated into pulsar population synthesis simulations. While \citet{Bagchi_odf} did provide software to compute the orbital degradation factors, the calculations (much like the search techniques themselves) are time-intensive and thus not optimized for inclusion in large scale population synthesis analysis such as \psrpoppy\ \citep{psrpoppy}.
    As a result, there has not been any significant modeling of the observed binary pulsar populations.
    
    In this work, we develop a computationally efficient framework to calculate the orbital degradation factor using the software provided by \citet{Bagchi_odf}. To do this, we integrate the orbital degradation factor into \psrpoppy\ (Sec.~\ref{sec_integration_w_poppy}), a pulsar population synthesis package designed to model the observed pulsar population discovered in multiple radio surveys at different radio frequencies \citep{psrpoppy}. We use this to place upper limits on the population of ultra-compact NS--WD and DNS systems in the Milky Way given that we have not yet detected any such system (Sec.~\ref{sec_ucb_pop}) and calculate the probability for any of the current large pulsar surveys to detect these UCB systems. Finally in  Sec.~\ref{sec_optimum_tint}, we calculate a range of optimum integration times that will maximize the S/N for UCB systems, thereby increasing the probability of detection of these systems in radio surveys.
    
\section{integrating orbital degradation factor into psrpoppy} \label{sec_integration_w_poppy}
    
    We use the framework developed in \citet{Bagchi_odf} to calculate the orbital degradation factor for a binary system. The orbital degradation factor, $\gamma$, can take values between $0 \leq \gamma \leq 1$ and when calculated at the harmonic, \m, depends on the mass of the pulsar, \ma, and the companion, \mb, the orbital period, \pb, eccentricity, \ecc, inclination, \inc, and angle of periastron passage, \om, as well as the spin period of the pulsar, \ps, and the integration time of the survey, \tobs\ \citep{og_odf}. The orbital degradation factor can be calculated for the case of a normal pulsar search (\gam) as well as pulsar searches which apply acceleration (\gama) and jerk (\gamj) search techniques. The radiometer S/N for the pulsar in the binary system is reduced by a factor of $\gamma_{i}^2$, where $i = [1,2,3]$ depending on the type of search technique. A lower orbital degradation factor implies a lower recovered S/N for the pulsar in the binary system due to Doppler smearing of the its signal from its orbital motion. We assume all modern pulsar surveys use acceleration search techniques and present results based on \gama\ in this work.
    
    The software to calculate the orbital degradation factor that was provided with \citet{Bagchi_odf} is computationally inefficient for use in large scale population synthesis simulations. To solve this problem, we used this software as a data generator to train a simple neural network to calculate the orbital degradation factor for a given binary system.
    
    \subsection{Data standardization} 
        The neural network takes the parameters described above as an input to calculate the orbital degradation factor. The range of training values for each of the input parameters are shown in Table~\ref{param_range}. These values are chosen to span the expected range of values for the population. We limit the harmonic, $m$, to a maximum value of 5, since the power in larger harmonics is relatively negligible for binary systems \citep{Bagchi_odf}. We also allow the companion mass, \mb, to vary up to $10^9$~M$_{\odot}$ to allow simulation of pulsars orbiting the Milky Way's central supermassive black hole, Sgr A*, which has a mass $\sim$10$^6$~M$_{\odot}$, as well as an even more massive hypothetical supermassive black hole. Since some of the input parameters can span multiple orders of magnitude, it is necessary to normalize the data to ease the training of the neural network. Thus, we first take the logarithm of the parameters \tobs, \ma, \mb, \ps, and \pb\ so that they have a dynamic range similar to the other input parameters. Next we normalize all of the input parameters such that they fall in the range between $\pm 1$.
        
        \begin{table}[]
            \centering
            \caption{In this table, we show the range of values of the input parameters for which the neural network presented in this work is trained.}
            \begin{tabular}{cccc}
                \toprule
                Name of parameter & units & Minimum & Maximum  \\
                \midrule
                Harmonic, \m & -- & 1 & 5 \\
                Survey integration time, \tobs\ & seconds & 1 & $5 \times 10^3$\\
                Mass of pulsar, \ma\ & $M_{\odot}$ & 1 & 2.4 \\
                Mass of companion, \mb\ & $M_{\odot}$ & 0.2 & $10^9$ \\
                Spin period of pulsar, \ps\ & seconds & $10^{-3}$ & 5 \\
                Inclination of binary system, \inc\ & degrees & $0^{\circ}$ & $90^{\circ}$ \\
                Angle of periastron passage, \om\ & degrees & $0^{\circ}$ & $360^{\circ}$ \\
                Eccentricity, \ecc & -- & 0 & 0.9 \\
                Orbital period, \pb\ & days & $10^{-3}$ & $10^{3}$ \\
                \bottomrule
            \end{tabular}
            \label{param_range}
        \end{table}
        
    \subsection{Network architecture}
        We use {\sc keras} \citep{keras_paper} with the {\sc TensorFlow} \citep{tf_paper} backend to develop our neural network model. 
        %We train a separate neural network for each of the three search techniques since the analytical equation for calculating the orbital degradation factor for each search technique is different \citep{Bagchi_odf}. 
        The neural network consists of five layers, with the input layer having 9 nodes (equal to number of inputs), three ``hidden'' layers containing 32 nodes, and the final, output layer consisting of a single node. We use the ``swish'' activation function \citep{swish_paper} for the hidden layers, while the output layer used a linear activation function. Since we assume that all surveys use the acceleration search technique, we will describe the training and performance of the neural network that models the acceleration search technique below (i.e.~\gama). However, the results are similar for the neural networks modeling the other search technique.
        
    \subsection{Training the neural network}
        We generate $69566$ combinations of the parameters described in Table~\ref{param_range} and calculate the corresponding \gama\ values using the software provided with \citet{Bagchi_odf}. 
        We take care to ensure that the training dataset spans the entire range of parameters described in Table~\ref{param_range}. We extract 5\% of this as a test dataset using which we can quantify the accuracy of the trained neural network. The remaining data have another 5\% reserved to be used as the validation dataset.
        
        During training, the neural network uses the input parameters to predict the orbital degradation factor (referred to as the prediction) which is then compared to the orbital degradation factor calculated using the analytical calculation (referred to as the label) in \citet{Bagchi_odf}. We use the mean absolute percentage error,
        \begin{equation}
            \displaystyle {\rm MAPE} = \left< 100 \times \frac{\left| \rm prediction - label \right|}{\rm label} \right>
        \end{equation}
        as the loss function for our neural network. We use the Adaptive Moment \citep[``ada'',][]{adam_paper} technique to optimize the learning for the neural networks and we stop training the neural network once the MAPE has stopped improving for the validation dataset. 
        
        \begin{figure}
            \centering
            \includegraphics[width = \columnwidth]{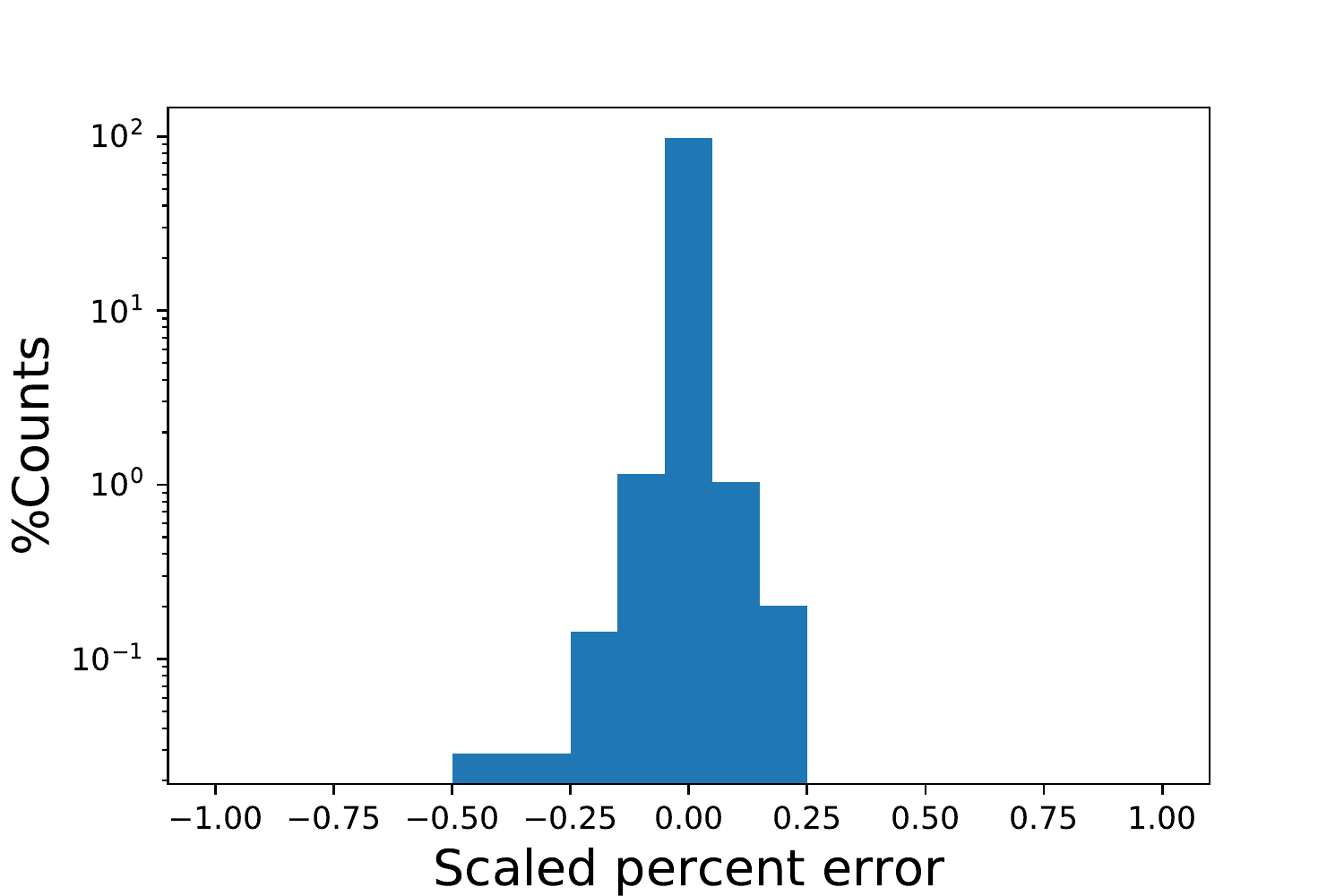}
            \caption{The distribution of the scaled percent error (Eq.~\ref{spe}) for the \gama\
            neural network evaluated on the training dataset.}
            \label{accuracy}
        \end{figure}
        
        The accuracy of this trained neural network can be calculated by evaluating its performance on the test dataset. We quantify the accuracy of the neural network through the distribution of the scaled percent error, 
        \begin{equation}
            \displaystyle \rm SPE = 100 \times \frac{\rm prediction - label}{label},
            \label{spe}
        \end{equation}
        which is shown in Fig.~\ref{accuracy}. 
        As we can see, 97.44\% of the predictions made by the neural network have an error of $\leq$5\% compared to the values predicted using the analytic solution from \citet{Bagchi_odf}, while there are almost no values with an error $\gtrsim$25\%. 
        
        The orbital degradation factor computation using the neural network framework is faster by a factor of $10^4$ compared to the same computation using the software provided by \citet{Bagchi_odf}. This demonstrates the suitability of the former for large scale population synthesis simulations. In addition, the inherent parallelism of the neural network framework allows it to compute multiple orbital degradation factors in a single pass while the software provided with \citet{Bagchi_odf} was limited to a single computation. This provides an additional significant improvement in the computational efficiency of the neural network framework.
        
\begin{figure*}
            \subfloat[]{\includegraphics[width = 0.49\textwidth]{"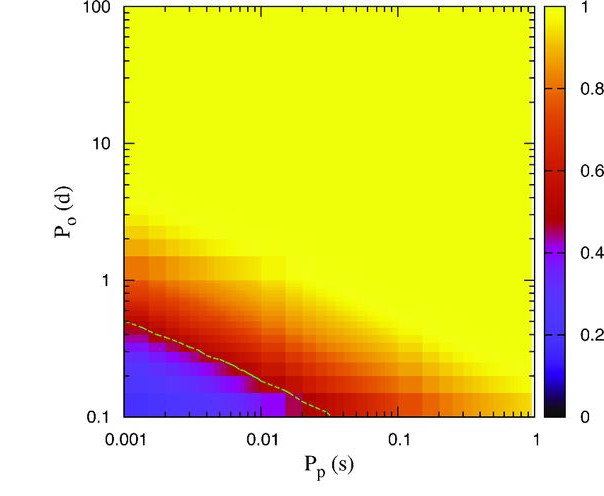"}}
            \hfill
            \subfloat[]{\includegraphics[width = 0.49\textwidth]{"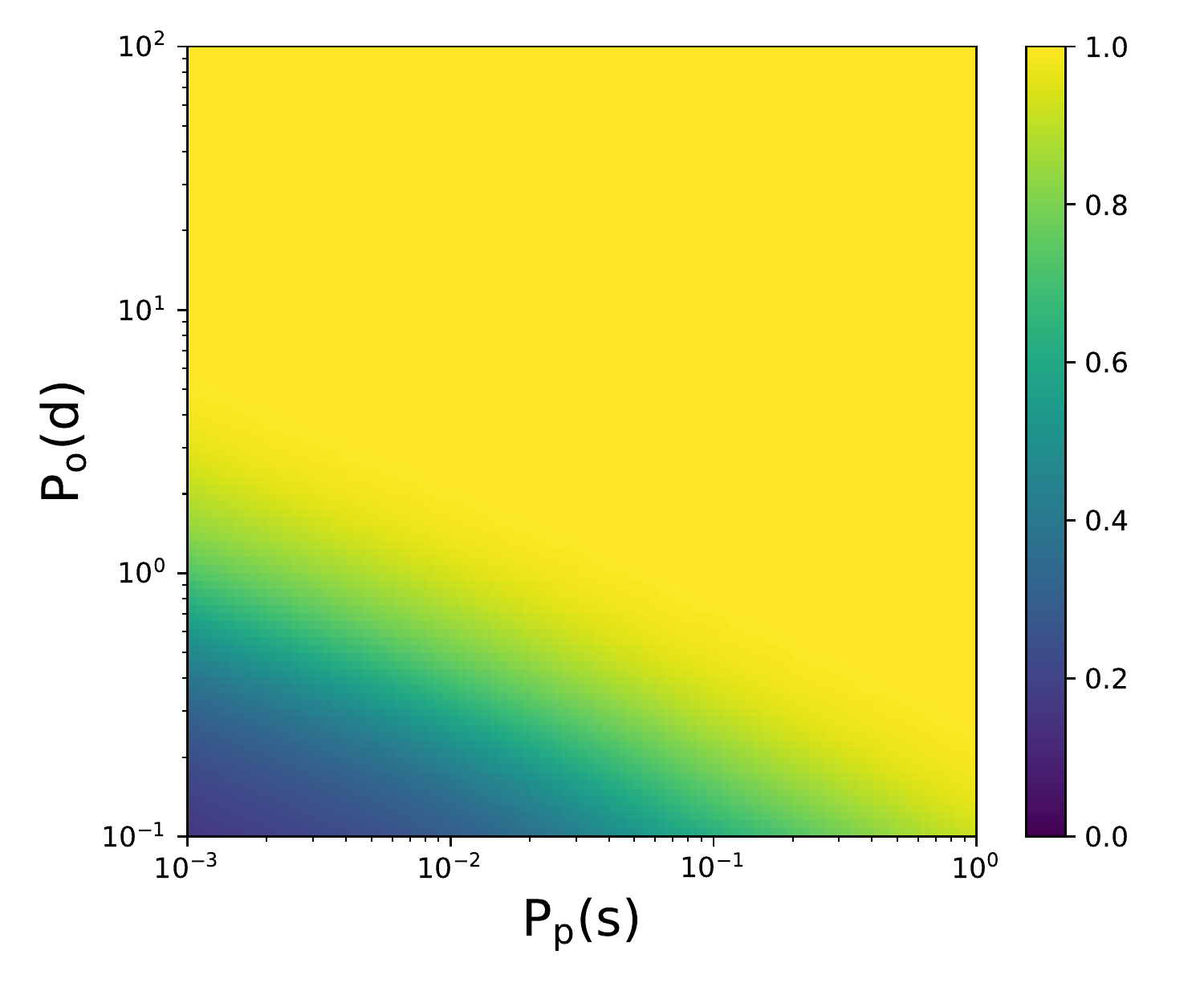"}}
            \caption{We directly compare the results from Fig. 9(b) in \citet{Bagchi_odf} (panel (a)) to the \gama\ neural network trained in this work (panel (b)). In both plots, the X-axis shows the spin period of the pulsar, the Y-axis shows the orbital period of the BNS system and the colorbar represents the orbital degradation factor. As we can see, the two methods produce identical results for the orbital degradation factor, \gama.}
            \label{fig_9b_comp}
        \end{figure*}

        We can directly compare the results produced by the trained neural network to those published in \citet{Bagchi_odf} by reproducing the figures from that work. As an example, in Fig.~\ref{fig_9b_comp}, we compare the results presented in Fig.~9(b) of \citet{Bagchi_odf} with those produced by our neural network. As we can see, the results produced by the two methods are identical, which is another confirmation of the accuracy of our neural network.
        
        %\begin{figure*}
        %    \centering
        %    \begin{subfigure}[b]{0.49\textwidth}
        %    \centering
        %    \includegraphics[width = \textwidth]{"fig9b.jpeg"}
        %    \caption{}
        %    \end{subfigure}
        %    \begin{subfigure}[b]{0.49\textwidth}
        %    \centering
        %    \includegraphics[width = %\textwidth]{"fig9b_nn.pdf"}
        %    \caption{}
        %    \end{subfigure}
        %    \caption{We directly compare the results from Fig. 9(b) in \citet{Bagchi_odf} (panel (a)) to the \gama\ neural network trained in this work (panel (b)). In both plots, the X-axis shows the spin period of the pulsar, the Y-axis shows the orbital period of the BNS system and the colorbar represents the orbital degradation factor. As we can see, the two methods produce identical results for the orbital degradation factor, \gama.}
        %    \label{fig_9b_comp}
        %\end{figure*}
        
    \subsection{Integration with \psrpoppy} \label{sec_integration}
        The orbital degradation factor calculated with the neural network can be directly integrated into \psrpoppy\ for modeling the different types of binary pulsar populations. 
        We add the ability for \psrpoppy\ to generate orbital parameters for a synthetic pulsar\footnote{\url{https://github.com/NihanPol/PsrPopPy2}} which are used to compute the orbital degradation factor.
        \psrpoppy\ calculates the S/N for a pulsar using the radiometer equation \citep{lorimer_kramer}, which can be directly scaled by $\gamma^2$ \citep{Bagchi_odf} to get the S/N for the same pulsar if it were in a binary. 
        
    \subsection{Selection bias against asymmetric mass DNS systems} \label{sec_asymm_bias}
        
        As an application of the orbital degradation factor, we investigate whether it is easier to detect a DNS system that is symmetric in mass as compared to the same system if it were asymmetric in mass.
        The question depends only on how the orbital degradation factor depends on the mass ratio of the binary system. To investigate this, we perform a Monte Carlo simulation where we randomly draw samples from the distributions for all the input parameters to the orbital degradation factor. 
        The majority of the observed sample of NS--WD and DNS systems have spin periods less than $\sim$100~ms \citep{psrcat}.
        Similarly, the majority of the observed NS--WD systems have orbital periods less than $\sim$50~days, while the majority of the observed DNS systems have orbital periods less than $\sim$10~days \citep{psrcat}.
        To correspond to the observed sample, we restrict the range of spin and orbital periods for the pulsars to be in the range $1 \, {\rm ms} < P_{\rm s} < 100 \, {\rm ms}$ and $10^{-3} \, {\rm days} < P_{\rm b} < 50 \, {\rm days}$ respectively. 
        Other parameters are allowed to vary across their full range as listed in Table~\ref{param_range}.
        However, in place of using the companion mass directly, we instead define a new parameter, the mass ratio, $q = m_1 / m_2$. We define symmetric systems as those having $0.9 \leq q \leq 1.0$ and asymmetric systems as $0.1 \leq q < 0.9$.
        
\begin{figure}
            \centering
            \includegraphics[width = \columnwidth]{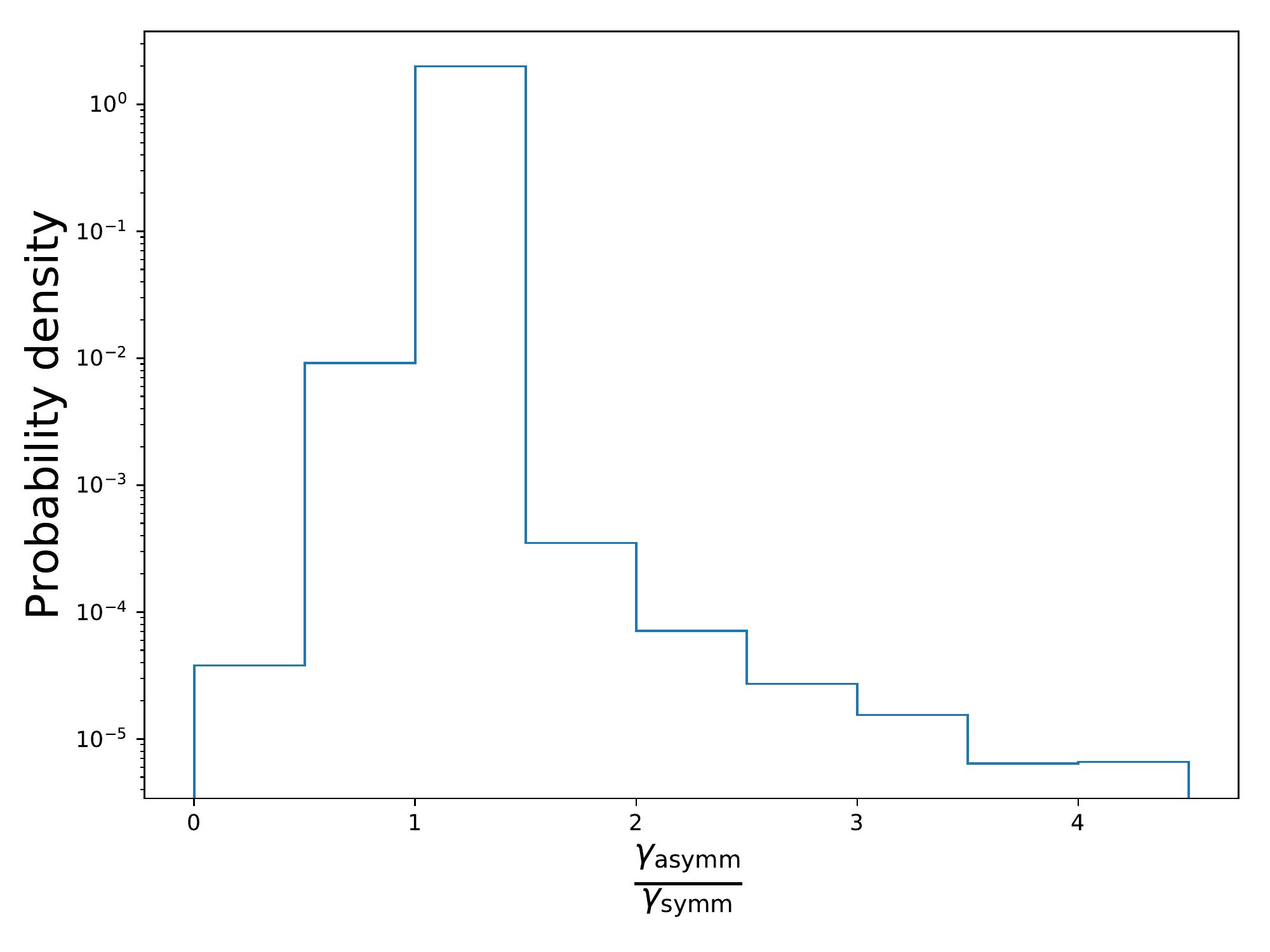}
            \caption{The ratio of the degradation factor for asymmetric mass systems to that of symmetric mass systems. The mass of the pulsar and all other orbital parameters, except the mass of the companion, are the same between the symmetric and asymmetric systems. The mass of the companion is calculated using the mass ratio, $q$, where for symmetric mass systems, $q \geq 0.9$ while for asymmetric mass systems, $q < 0.9$. The histogram shows that it is easier to detect asymmetric mass systems that symmetric mass systems.}
            \label{asymm_pref}
        \end{figure}

        We randomly draw a value for the mass ratio for symmetric and asymmetric systems as defined above. Both of these mass ratio values are then assigned the same set of remaining input parameters required to calculate the orbital degradation factor. We then calculate and plot the distribution of the ratio of the orbital degradation factor for asymmetric systems to that for symmetric systems. The distribution obtained after $10^7$ sample draws is shown in Fig.~\ref{asymm_pref}. 
        
        As we can see there are fewer systems in which the ratio has a value less than one, implying that the orbital degradation factor for asymmetric DNS systems is on average greater than that for a symmetric DNS systems. Consequently, asymmetric DNS systems are easier to detect in surveys as compared to symmetric DNS systems.
        However, despite the preference for the detection of asymmetric mass DNS systems, only two such systems have been detected, J0453+1559 \citep[$q = $][]{most_asymm_bns_0453} and J1913+1102 \citep[$q = 0.75$][]{my_nature_paper}, compared to eighteen other DNS systems with mass ratios $q \gtrsim 0.9$. This result suggests that this discrepancy in the number of detected asymmetric systems might not be due to selection effects, but rather due to differences in the evolutionary scenarios between the two types of systems.
        
\section{Ultra-compact binary population statistics} \label{sec_ucb_pop}
    
    \subsection{Size of population} \label{subsec_pop_num}
        
        \begin{table*}
            \centering
            \caption{This table lists the telescope and survey parameters for the large pulsar surveys that are considered in this work.}
            \begin{tabular}{ccccccc}
                \toprule
                Survey & Gain, $G$ & Center Frequency, $f_{\rm c}$ & Bandwidth, $B$ & System temperature, $T_{\rm sys}$ & Integration time, $t_{\rm int}$ \\
                -- & (K/Jy) & (MHz) & (MHz) & (K) & (s) \\
                \midrule
                PALFA\footnote{Pulsar Arecibo L-band Feed Array, \citet{PALFA}} & 8.5 & 1374 & 300 & 25 & 268 \\
                PMSURV\footnote{Parkes Multibeam SURvey, \citet{PMSURV}} & 0.6 & 1374 & 288 & 25 & 2100 \\
                AODRIFT\footnote{ArecibO DRIFT scan survey, \citet{aodrift_1}} & 10 & 327 & 25 & 100 & 50 \\
                GBNCC\footnote{Green Bank North Celestial Cap Survey, \citet{gbncc}} & 2 & 350 & 100 & 46 & 120 \\
                HTRU--low\footnote{High Time Resolution Universe low-latitude survey \citet{htru_low_mid}} & 0.6 & 1352 & 340 & 25 & 340 \\
                HTRU--mid\footnote{High Time Resolution Universe mid-latitude survey \citet{htru_low_mid}} & 0.6 & 1352 & 340 & 25 & 540 \\
                \bottomrule
            \end{tabular}
            \label{survey_table}
        \end{table*}
        
        Given the non-detection of UCB systems by current radio pulsar surveys, we can place an upper limit on the number of these systems in the Galaxy. To do so, we use the version of \psrpoppy  \citep{psrpoppy} that is integrated with the orbital degradation factor described in Sec.~\ref{sec_integration}. We compute separately the upper limit on the population of ultra-compact NS--WD and DNS systems.
        
        We follow a procedure that is based on the framework described in \citet{kkl}. For any given type of binary system, if $N_{\rm obs}$ is the number of observed systems, we expect its probability distribution  to follow a Poisson distribution:
        \begin{equation}
            \displaystyle P(N_{\rm obs}; \lambda) = \frac{\lambda^{N_{\rm obs}} e^{-\lambda}}{N_{\rm obs}!}
            \label{poisson}
        \end{equation}
        where, by definition, $\lambda = \left< N_{\rm obs} \right>$. As described in \citet{kkl}, we know that the relation
        \begin{equation}
            \displaystyle \lambda = \alpha N_{\rm tot}
            \label{hypo}
        \end{equation}
        is true, where $N_{\rm tot}$ is the number of UCB pulsars that are beaming towards the Earth and $\alpha$ is a constant that depends on the properties of the UCB system and the pulsar surveys under consideration. Since no UCB systems have been detected, we can set $N_{\rm obs} = 0$, which reduces Eq.~\ref{poisson} to $P(0; \lambda) = e^{-\lambda}$.
        
        As described in \citet{kkl}, the likelihood function, $P(D|HX)$, where $D = 0$ is the real observed sample, $H$ is our model hypothesis (i.e. Eq~\ref{hypo}), and X is the population model, is defined as,
        \begin{equation}
            \displaystyle P(D|HX) = P(0|\lambda(N_{\rm tot}), X) = e^{-\lambda(N_{\rm tot})}.
            \label{likely}
        \end{equation}
        Using Bayes' theorem and the justification given in \citet{kkl}, the posterior, $P(\lambda|DX)$, is equal to the likelihood function, i.e.,
        \begin{equation}
            \displaystyle P(\lambda|DX) \equiv P(\lambda) = P(0| \lambda, X) = e^{-\lambda(N_{\rm tot})}.
            \label{posterior}
        \end{equation}
        Using this posterior, we can calculate the probability distribution for $N_{\rm tot}$,
        \begin{equation}
            \displaystyle P(N_{\rm tot}) = P(\lambda) \left| \frac{d\lambda}{dN_{\rm tot}} \right| = \alpha e^{- \alpha N_{\rm tot}}.
            \label{pop_prob}
        \end{equation}
        
        With {\psrpoppy}, we generate populations of different sizes and calculate $\lambda$ for each population using Eq.~\ref{poisson}, which in combination with Eq.~\ref{hypo}, gives us the value of $\alpha$. Using Eq.~\ref{pop_prob} with the value of $\alpha$ gives us the probability density for the population of the UCB systems that are beaming towards Earth.
        
        For all UCB systems, we allow the mass of the pulsar, \ma, inclination of the system, \inc, the angle of periastron passage, \om, and the eccentricity, \ecc, to have the range listed in Table~\ref{param_range}. For ultra-compact NS--WD systems, we restrict the companion mass to the range $0.2 \, M_{\odot} < m_2 < 1.4 \, M_{\odot}$, while for ultra-compact DNS systems, we restrict the companion mass to the range $1.0 \, M_{\odot} < m_2 < 2.4 \, M_{\odot}$. We also assume the pulsar is an orthogonal rotator and thus, most of the power from the pulsar emission is constrained in the second harmonic, and set $m = 2$. In the case that the pulsar is not an orthogonal rotator, a choice of $m = 2$ results in a more conservative upper limit. For both ultra-compact NS--WD and DNS systems, we constrain the orbital period to the range $1.5 \, {\rm minutes} < P_{\rm b} < 15 \, {\rm minutes}$. All of these parameters have uniform distributions.
        
        In addition, we constrain the spin period for DNS systems to the range $1 \, {\rm ms} < P_{\rm s} < 100 \, {\rm ms}$ to correspond to the observed spin periods distribution and for NS--WD systems we assume a log-normal period distribution with mean and standard deviation of 4.5~ms and 1.8~ms respectively \citep{dunc_nswd_pdist}. We model the pulsar luminosity distribution using a log-normal distribution with a mean $\left< log_{10}L \right> = -1.1$ ($L = 0.07 \, {\rm mJy \, kpc^2}$) and standard deviation $\sigma_{\rm log_{10}L} = 0.9$ \citep{fk06}. Since we consider surveys at different radio frequencies, we also model the pulsar spectral index as a normal distribution with mean $\alpha = -1.4$ and standard deviation $\beta = 1$ \citep{bates_si_dist}. We assume the radial distribution for the UCB systems as described in \citet{lorimer_rad_dist} and the two-sided exponential function for the $z$-height distribution, with a scale height of $z_0 = 0.33$~kpc.
        
\begin{figure*}
            \centering
            \subfloat[]{\includegraphics[width = 0.49\textwidth]{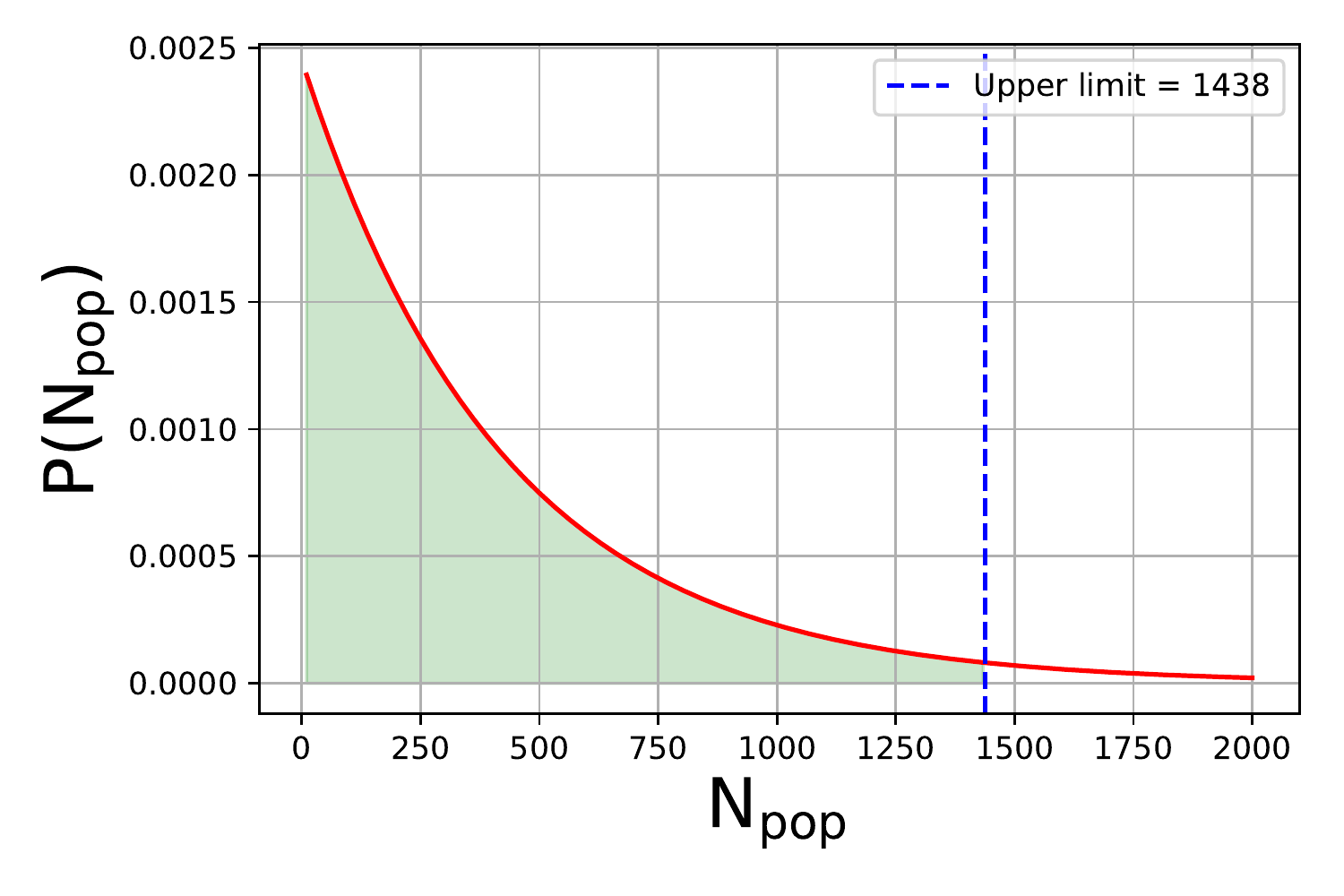}}
            \hfill
            \subfloat[]{\includegraphics[width = 0.49\textwidth]{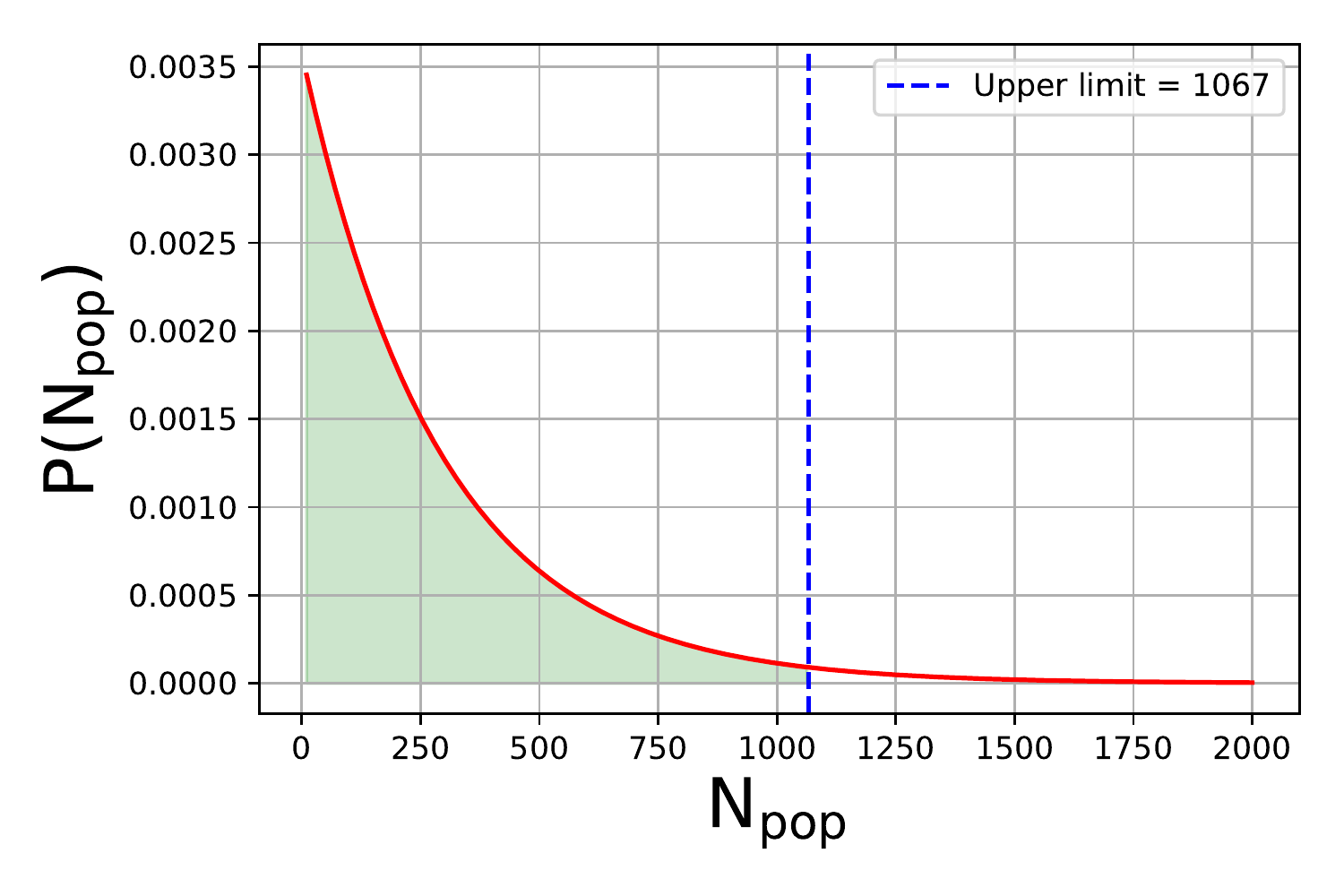}}
            \caption{Probability distribution function for the number of UCB systems that are beaming towards the Earth. As we can see, the 95\% upper limit on the number of ultra-compact NS--WD systems (panel (a)) is 1450 systems, while that for ultra-compact DNS systems (panel (b)) is 1100 systems.}
            \label{pops}
        \end{figure*}

        Finally, the surveys that we consider are listed in Table~\ref{survey_table}. These are the largest radio pulsar surveys conducted to date. The integration times from the individual surveys are used in the calculation of the orbital degradation factor. Using these parameters, the probability distribution for the size of the population of the UCB systems that are beaming towards the Earth is shown in Fig.~\ref{pops}. As we can see, the 95\% upper limit on the number of ultra-compact NS--WD systems in the Galaxy is 1450 systems, while that for ultra-compact DNS systems in the Galaxy is 1100 systems.
        
        %\begin{figure*}
        %    \centering
        %    \begin{subfigure}[b]{0.49\textwidth}
        %        \centering
        %        \includegraphics[width = \textwidth]{"nswd_npop".pdf}
        %        \caption{Ultra-compact NS--WD systems}
        %    \end{subfigure}
        %    \begin{subfigure}[b]{0.49\textwidth}
        %        \centering
        %        \includegraphics[width = \textwidth]{"dns_npop".pdf}
        %        \caption{Ultra-compact DNS systems}
        %    \end{subfigure}
        %    \caption{Probability distribution function for the number of UCB systems that are beaming towards the Earth. As we can see, the 95\% upper limit on the number of ultra-compact NS--WD systems (panel (a)) is 1450 systems, while that for ultra-compact DNS systems (panel (b)) is 1100 systems.}
        %    \label{pops}
        %\end{figure*}
        
        %The upper limit on the number of ultra-compact NS--WD systems is slightly smaller than that of ultra-compact DNS systems due to the fact that the orbital degradation factor for the former is on average larger than that for the latter. Consequently, ultra-compact NS--WD systems are easier to detect than ultra-compact DNS system and thus their non-detection so far places a more stringent constraint on their population as compared to ultra-compact DNS systems. Note that this does not necessarily imply that there are fewer ultra-compact NS--WD binaries than DNS binaries, but merely states that we can constrain the population of the former type of system better than that of the latter.
        
        As stated earlier, the upper limits above are the number of UCB systems that are beaming towards Earth. The total number of these systems in the Galaxy can be calculated by scaling $N_{\rm tot}$ by the beaming correction factor, $f_{\rm b}$ \citep{kkl}. Given the large uncertainty in the beaming correction factors, if we use the average beaming correction factor measured for the merging DNS systems, $f_{\rm b} = 4.6$ \citep{my_merger_rate}, the upper limit on the total number of ultra-compact NS--WD and DNS systems comes out to $\sim$6700 and $\sim$5000 systems respectively.
        
        The number of ultra-compact DNS systems derived here is less than the total number of merging DNS systems derived in \citet{my_merger_rate} and \citet{my_nature_paper}. This difference can be explained by the fact that the UCB systems that we consider in this work have lifetimes $\sim$ few Myr, significantly smaller than that for the merging DNS systems studied in the aforementioned studies. As a result, these systems are closer to merger and spend a relatively short amount of time in this subclass of DNS systems compared to the larger orbital period merging DNS systems from \citet{my_merger_rate}, which results in an overall smaller population size of ultra-compact DNS systems. The upper limit on the number of ultra-compact DNS systems is also consistent with recent estimates of the size of this population made by \citet{lau_detecting_dns_w_lisa} and \citet{lisa_dns_andrews}. Note that in these simulations, we assumed that the luminosity function for both of these types of systems is the same and is the same as that of the observed pulsar population \citep{fk06}. However, if the luminosity function for these systems is different, that will also influence the number of such systems in the galaxy.
        
    \subsection{Probability of pulsar surveys detecting an UCB system} \label{subsec_survey_eff}
        
        Knowing the upper limit on the number of UCB systems that are beaming towards us, we can calculate the probability of the radio pulsar surveys listed in Table~\ref{survey_table} in detecting these systems. To do so, we assume that the number of the UCB systems (both NS--WD and DNS) in the Galaxy is equal to their upper limits, i.e. we calculate the probability for these surveys in the most optimistic scenario.
        
\begin{figure*}
            \centering
            \subfloat[]{\includegraphics[width = 0.49\textwidth]{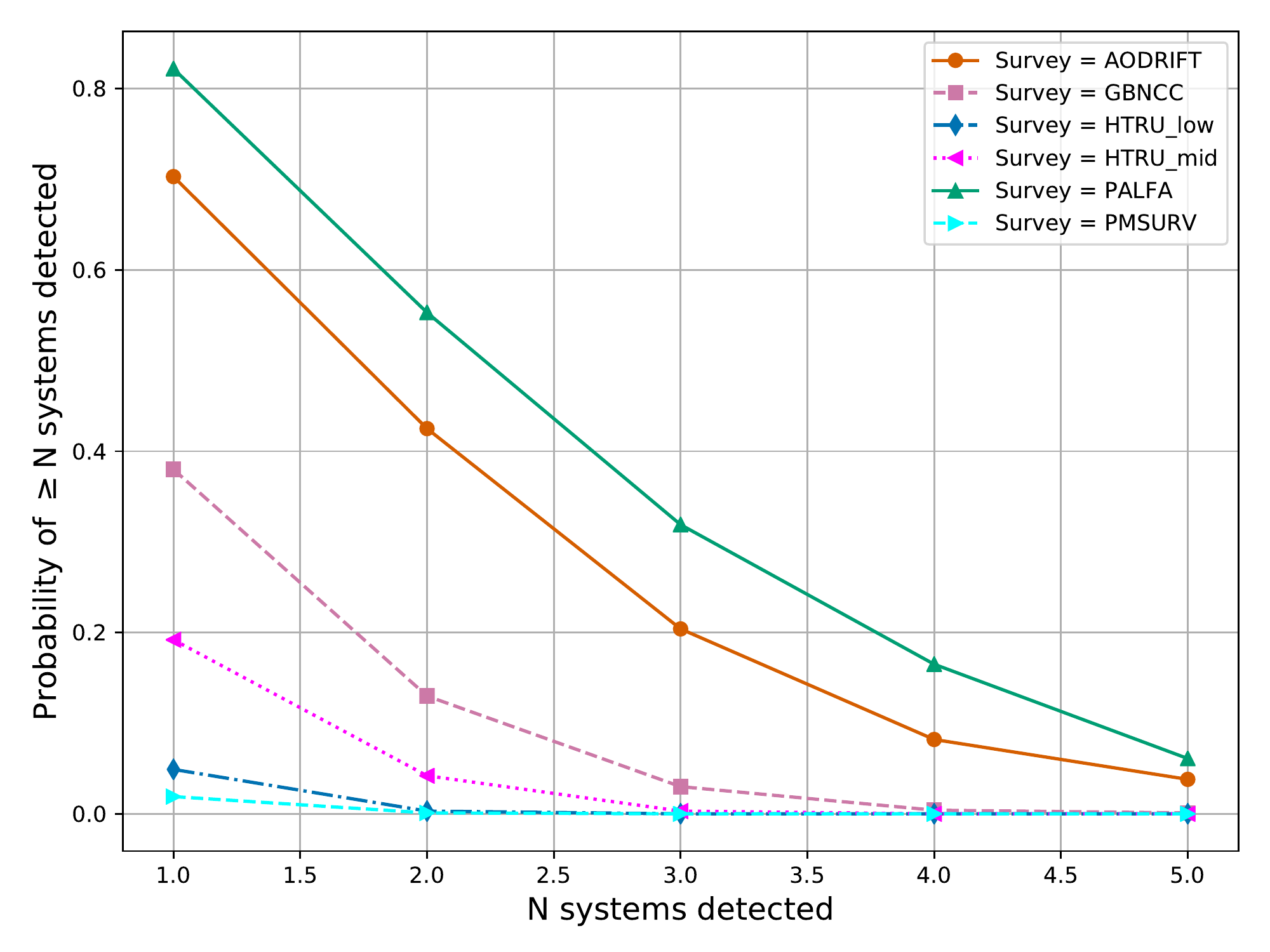}}
            \hfill
            \subfloat[]{\includegraphics[width = 0.49\textwidth]{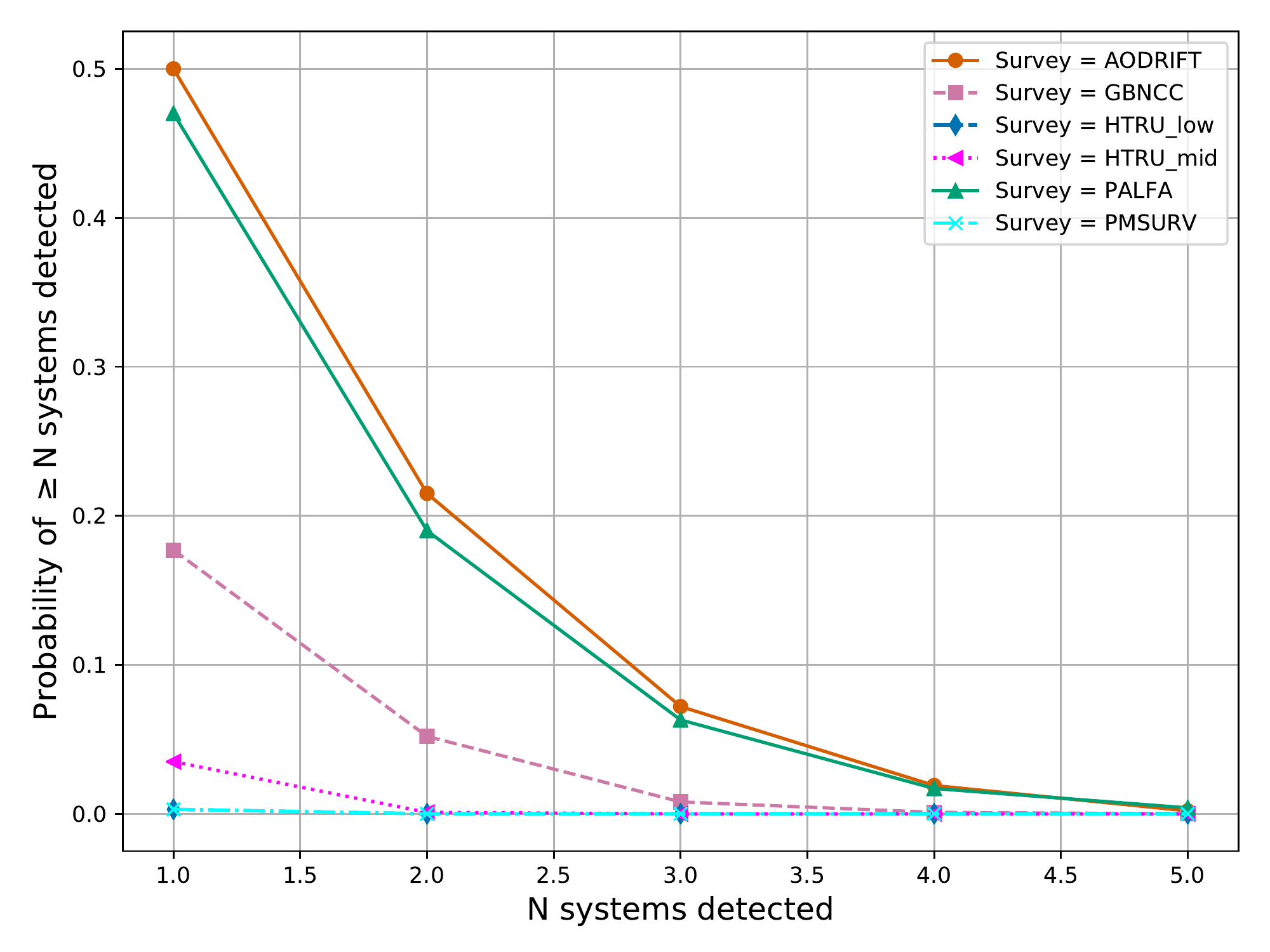}}
            \caption{Complementary cumulative distribution function of the number of UCB systems that are detectable by the radio pulsar surveys listed in Table~\ref{survey_table}. The X-axis shows the number of detected systems, $N$, while the Y-axis shows the probability that $\geq$N systems will be detected in the given survey. The surveys conducted with the Arecibo telescope, i.e. PALFA and AODRIFT, are most likely to detect at least one of these UCB systems.}
            \label{success_survey}
        \end{figure*}

        Next, we use \psrpoppy\ to generate this number of pulsars in the Galaxy, with the orbital, spin, luminosity, spatial and spectral index distributions being the same as described in Sec.~\ref{subsec_pop_num}. We generate $10^3$ different versions of these populations to ensure that we are efficiently sampling all of the prior distributions. Accounting for the orbital degradation factor, we then ``run'' each of the surveys listed in Table~\ref{survey_table} on each of these populations and count the number of systems that are detected by each survey. We then calculate the complementary cumulative distribution function for the number of detections by each survey, which is shown in Fig.~\ref{success_survey}. 
        
        %\begin{figure*}
        %    \centering
        %    \begin{subfigure}[b]{0.49\textwidth}
        %        \centering
        %        \includegraphics[width = \textwidth]{"success_nswd".pdf}
        %        \caption{Ultra-compact NS--WD systems}
        %    \end{subfigure}
        %    \begin{subfigure}[b]{0.49\textwidth}
        %        \centering
        %        \includegraphics[width = \textwidth]{"success_dns".pdf}
        %        \caption{Ultra-compact DNS systems}
        %    \end{subfigure}
        %    \caption{Complementary cumulative distribution function of the number of UCB systems that are detectable by the radio pulsar surveys listed in Table~\ref{survey_table}. The X-axis shows the number of detected systems, $N$, while the Y-axis shows the probability that $\geq$N systems will be detected in the given survey. The surveys conducted with the Arecibo telescope, i.e. PALFA and AODRIFT, are most likely to detect at least one of these UCB systems.}
        %    \label{success_survey}
        %\end{figure*}
        
        As we can see, the surveys with the Arecibo radio telescope, i.e. PALFA and AODRIFT, have the highest probability to detect $\geq$1 of these UCB systems. This is followed by the GBNCC and HTRU mid-latitude survey, while the HTRU low-latitude and PMSURV have the least probability of detecting any of these UCB systems.
        
        The difference in the efficiency of these surveys at detecting the UCB systems is due to the integration times used for processing these surveys. As can be seen in Table~\ref{survey_table}, AODRIFT has the shortest integration time of all the surveys, followed by GBNCC, PALFA, and the HTRU mid-latitude survey. A shorter integration time always produces a larger degradation factor, thereby providing a larger S/N detection for these systems.
        
        However, it is possible to use a longer integration time and still maintain sensitivity to UCB systems, as is demonstrated by the PALFA survey. While the PALFA survey has the third-shortest integration time in Table~\ref{survey_table}, it is able to offset the relative loss in S/N due to the orbital degradation by increasing the overall sensitivity of the radio telescope. We discuss the balancing of the survey integration time with the orbital degradation factor further in Sec.~\ref{sec_optimum_tint}.
        
    \subsection{Multi-messenger prospects for detectable ultra-compact binaries}
        
        In this optimistic scenario where the number of UCB systems beaming towards the Earth corresponds to the 95\% upper limit derived above, we can expect to detect as many as four of these UCB systems with the radio pulsar surveys at Arecibo alone. Given their short orbital periods, these UCB systems could be promising sources for LISA, the space-based gravitational wave observatory that is scheduled to launch in the 2030s \citep{LISA}. To see if these systems will be detectable with LISA, we need to compute the S/N for these systems with respect to LISA's sensitivity curve \citep{lisa_sensitivity_curve}.
        
        For a binary system with eccentricity, \ecc, the GW emission from the system is spread over multiple harmonics of the orbital frequency, $f_{\rm n} = n / P_{\rm b}$, where $n$ represents the n'th harmonic. The total S/N for these systems as observed by LISA can be calculated as the quadrature sum of the S/N at each of these harmonics \citep{lisa_ecc_strain_1, lisa_ecc_strain_2},
        \begin{equation}
            \displaystyle {\rm S/N^2} \approx \sum_{n = 1}^{\infty} \frac{h_{\rm n}^2(f_{\rm n}) T_{\rm LISA}}{S_{\rm LISA}(f_{\rm n})}
            \label{lisa_snr}
        \end{equation}
        where $S_{\rm LISA}(f_{\rm n})$ is the LISA sensitivity curve as defined by \citet{lisa_sensitivity_curve}, $T_{\rm LISA} = 4$~yrs is the timespan of the LISA mission, and $h_{\rm n}(f_{\rm n})$ is the strain amplitude,
        \begin{equation}
            \displaystyle h_{\rm n}(f_{\rm n}) = \frac{8}{\sqrt{5}} \left( \frac{2}{n} \right)^{5/3} \frac{(\pi f_{\rm n})^{2/3} (\mathcal{G} \mathcal{M})^{5/3}}{c^4 d} \sqrt{g(n, e)}
            \label{lisa_strain}
        \end{equation}
        where $\mathcal{G}$ is the Gravitational constant, $c$ is the speed of light, $d$ is the distance to the binary system, $\mathcal{M} = m_1^{3/5} m_2^{3/5} (m_1 + m_2)^{-1/5}$ is the chirp mass of the binary, and $g(n, e)$ provides the relative amplitude between the different harmonics \citep[see Eq.~20 in][]{peters_mathews_1963}.
        
        Using these relations, we can calculate the S/N with which LISA would observe the UCB systems that are detected with the radio pulsar surveys described above. For the UCB binaries that were detected in the simulations described in Sec.~\ref{subsec_survey_eff} (both NS--WD and DNS), we extract the mass, \ma\ and \mb, of the components of the UCB system, the orbital period, \pb, eccentricity, \ecc, and radial distance to the UCB system, $d$. We remind the reader that the radial distribution of the pulsars in the Galaxy was assumed to be the one described in \citet{lorimer_rad_dist}, while the $z$-height distribution was the one described in \citet{lyne_z_dist}. Given these parameters, we calculate the strain using Eq.~\ref{lisa_strain} at harmonics $2 \leq n \leq 30$ and then calculate the S/N for each system by summing over these harmonics as described in Eq.~\ref{lisa_snr}.
        
        All of the UCB systems that were detected in the radio pulsar surveys have an S/N that is comfortably above the LISA threshold $\rm S/N = 7$ assuming a four year LISA mission. 
        %Thus, even if the UCB systems have the same spatial distribution as the observed pulsars in the Galaxy, that will not preclude them from being detectable by LISA.
        Thus, as shown in Sec.~\ref{subsec_survey_eff}, even if radio pulsar surveys are able to detect only a couple of these UCB systems, these should be strong detections for LISA and will allow for multi-messenger studies of neutron stars \citep[for example, see][]{thrane_lisa_dns}. 
        
\section{Optimum integration time for detecting ultra-compact BNS systems} \label{sec_optimum_tint}
    
    As stated earlier, the impact of the orbital motion of the pulsar on the S/N is most acutely felt when the pulsar is part of an ultra-compact binary system (UCB).
    The modified radiometer equation for pulsars, including the orbital degradation factor, $\gamma$, can be written as \citep{lorimer_kramer}:
    \begin{multline}
        \displaystyle {\rm S/N} = \left[ \frac{G \sqrt{B N_{\rm p}}}{T_{\rm sys}} \right] \, S \, \left[ \sqrt{\frac{t_{\rm int} (P_{\rm s} - w)}{w}} \right] \, \gamma(t_{\rm int}, P_{\rm s}, ...)^2 \\
        = \xi \times S \times f_1(t_{\rm int}, P_{\rm s}, w) \times \gamma(t_{\rm int}, P_{\rm s},...)^2 
        \label{radiometer_eq}
    \end{multline}
    where $S$ is the pulsar flux, $\xi$ is a constant that depends on the telescope and survey setup such as $G$, the receiver gain, $B$, the receiver bandwidth, $N_{\rm p} = 2$, the number of polarizations, $T_{\rm sys}$, the receiver system temperature (which includes the sky temperature, $T_{\rm sky}$), $P_{\rm s}$ is the spin period of the pulsar and $w$ is the effective pulse width (which includes effects of dispersion smearing and scattering). We list the telescope and survey parameters, as well as the integration times for the large pulsar surveys that we analyze in this work in Table~\ref{survey_table}.
    
    \begin{figure}
        \centering
        \includegraphics[width = \columnwidth]{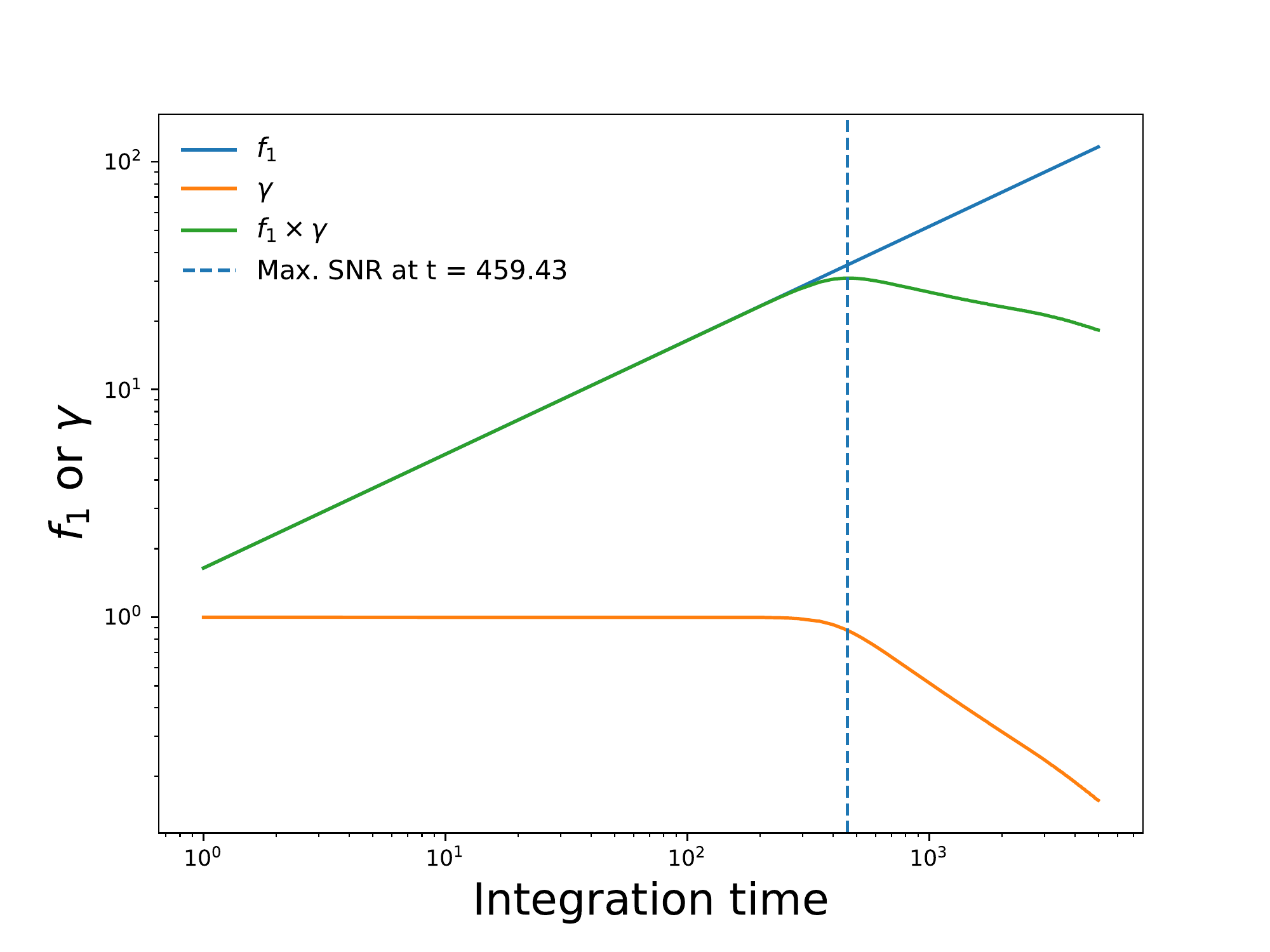}
        \caption{An example of maximizing the S/N by optimizing the selection of the integration time for the pulsar survey. This example uses the spin and orbital parameters of J0737--3039A \citep[Double Pulsar,][]{0737A_disc}. The function, $f_1$ (Eq.~\ref{radiometer_eq}), is directly proportional to the integration time, while there is a ``knee" in the orbital degradation factor.
        As a result, the maximum S/N for this pulsar would be at an integration time of $t_{\rm int} = 459$~s.}
        \label{motivate_opt_tint}
    \end{figure}
    
    For a pulsar with flux, $S$, we need to select an integration time, $t_{\rm int}$, that maximizes the observed S/N for the pulsar in the BNS system. To motivate the problem, we show in Fig.~\ref{motivate_opt_tint} an example for a system that has the orbital and spin properties of the millisecond pulsar in the Double Pulsar system
    %i.e. $P_{\rm s} = 22.7\,{\rm ms}; \ P_{\rm b} \approx 0.1\,{\rm days}; \ e = 0.08; \ \omega_{\rm p} = 179.98$
    \citep{Lyne_dpsr, 0737A_disc}. As we can see, when the orbital degradation factor is not considered in the S/N calculation, the S/N for the system grows as a function of the integration time, $\rm S/N \propto \sqrt{t_{\rm int}}$. However, the orbital degradation factor function for this system has a ``knee'' at an integration time of $\sim$460~s, after which the orbital degradation factor decreases with increasing integration time. As a result of this ``knee'' feature, the total S/N for the system peaks at the position of the ``knee'' introduced by the orbital degradation factor. Thus, the integration time corresponding to this peak would be the optimum integration time to detect systems like the Double Pulsar. Similarly, each binary pulsar system will have its own unique optimum integration time.
    
    Note that $\xi$ from Eq.~\ref{radiometer_eq} does not affect the optimum integration time, but will affect the final S/N of the system. This factor encodes the instrumental sensitivity of the telescope that is used for a given pulsar survey and thus, this factor will be larger for a more sensitive telescope. For example, for the PALFA survey at Arecibo, $\xi = 8.33$, while for the GBNCC survey at the Green Bank Telescope, $\xi = 0.61$. In fact, PALFA has the largest $\xi$ value for any survey listed in Table~\ref{survey_table}, which explains why it has a high probability of detecting an UCB system despite having the third-shortest integration time (see Sec.~\ref{subsec_survey_eff}). Despite this, it is still important to derive and use an optimum integration time to maximize the probability of detecting an UCB system with all the surveys.
    
    To generalize the example described above, we use a Monte Carlo simulation similar to the one described in Sec.~\ref{sec_asymm_bias}. Since we are interested in UCB systems, we constrain the mass of the companion and the orbital period to the range $0.2~{\rm M}_{\odot} < m_2 < 2.4~{\rm M}_{\odot}$ and $1.5 \, {\rm min} < P_{\rm b} < 15 \,$min. 
    %We also constrain the spin periods to the range $1 \, \rm ms < P_{\rm s} < 100 \, \rm ms$ to correspond to the periods seen for the BNS systems in the Galaxy (see Sec.~\ref{sec_asymm_bias}), 
    In this case, as we desire simply a range of optimal integration times, we sample a uniform distribution of spin periods in the range 1 to 100 ms
    and assume a fixed duty cycle of $\delta = 0.06$ \citep{lorimer_kramer} and fix the harmonic to $m = 2$ (see Sec.~\ref{subsec_pop_num}). The other parameters have the same range as listed in Table~\ref{param_range}. We draw $10^7$ random samples from these distributions and calculate the optimum integration time as described above for each UCB system.
    
    The above analysis yields an optimum integration time of $t_{\rm opt} = 42^{+153}_{-22}$~s, where the errors represent the 95\% confidence intervals on the peak of the distribution. Comparing this time to the integration times used for the large pulsar surveys in Table~\ref{survey_table}, we can see that the AODRIFT and GBNCC surveys are ideally placed towards detecting UCB systems, while PALFA is able to compensate for the loss in S/N by having a high $\xi$ value as described above. This is also seen in Fig.~\ref{success_survey}, where these three surveys have the highest probability of detecting at least one UCB system.
    
    %\begin{figure*}
    %        \centering
    %        \begin{subfigure}[b]{0.49\textwidth}
    %            \centering
    %            \includegraphics[width = \textwidth]{"ideal_success_nswd".pdf}
    %            \caption{Ultra-compact NS--WD systems}
    %        \end{subfigure}
    %        \begin{subfigure}[b]{0.49\textwidth}
    %            \centering
    %            \includegraphics[width = \textwidth]{"ideal_success_dns".pdf}
    %            \caption{Ultra-compact DNS systems}
    %        \end{subfigure}
    %        \caption{Complementary cumulative distribution function of the number of UCB systems that are detectable by the radio pulsar surveys listed in Table~\ref{survey_table}, but with integration time set to the optimum value of $\sim$50~s derived in Sec.~\ref{sec_optimum_tint}. The X-axis shows the number of detectable systems, $N$, while the Y-axis shows the probability that $\geq$N systems will be detected in the given survey. Compared to Fig.~\ref{success_survey}, we can see a significant improvement in the detection probability for all the surveys.}
    %        \label{ideal_success_survey}
    %\end{figure*}
    \begin{figure*}
        \centering
        \subfloat[]{\includegraphics[width = 0.49\textwidth]{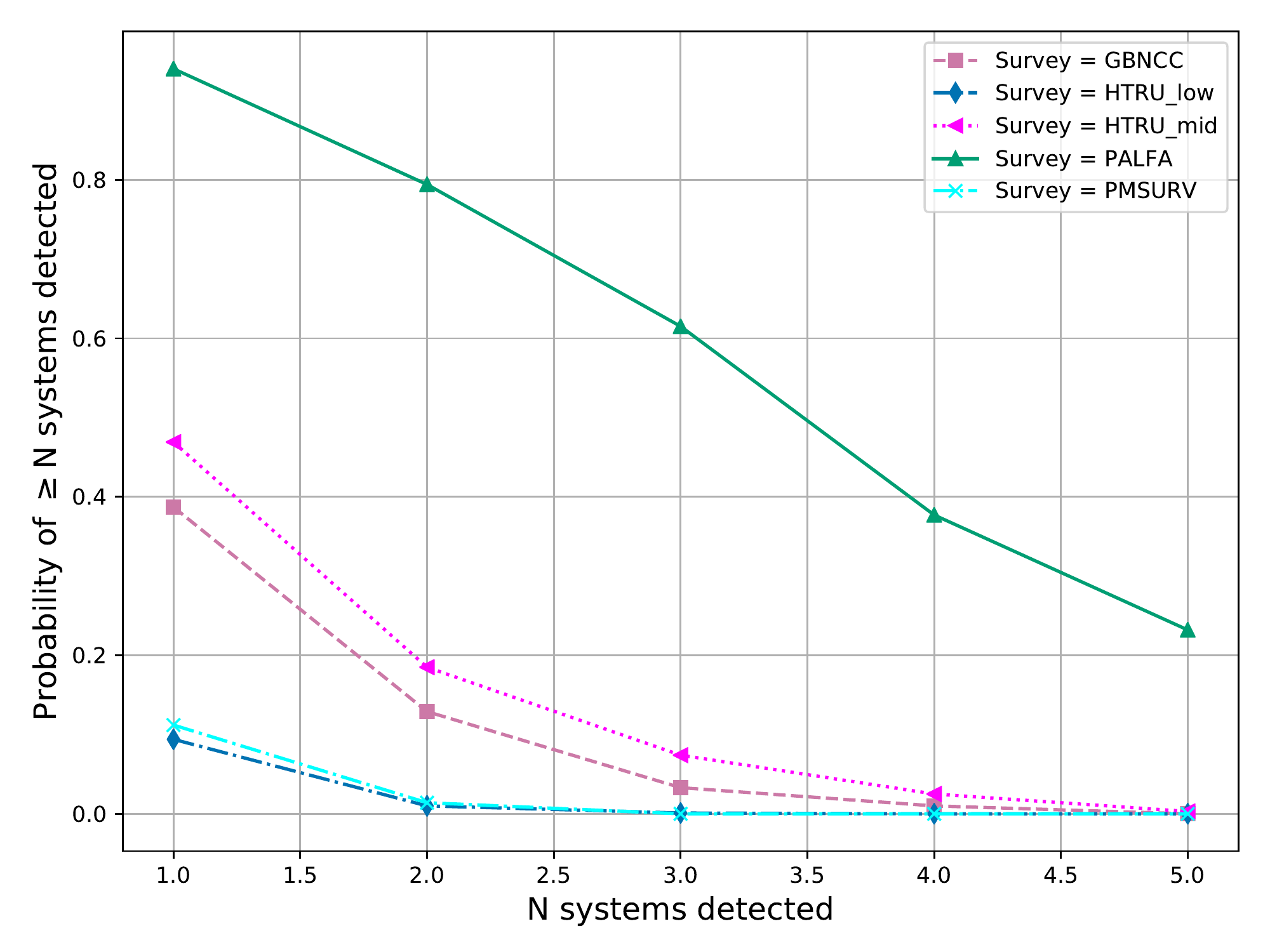}}
        \hfill
        \subfloat[]{\includegraphics[width = 0.49\textwidth]{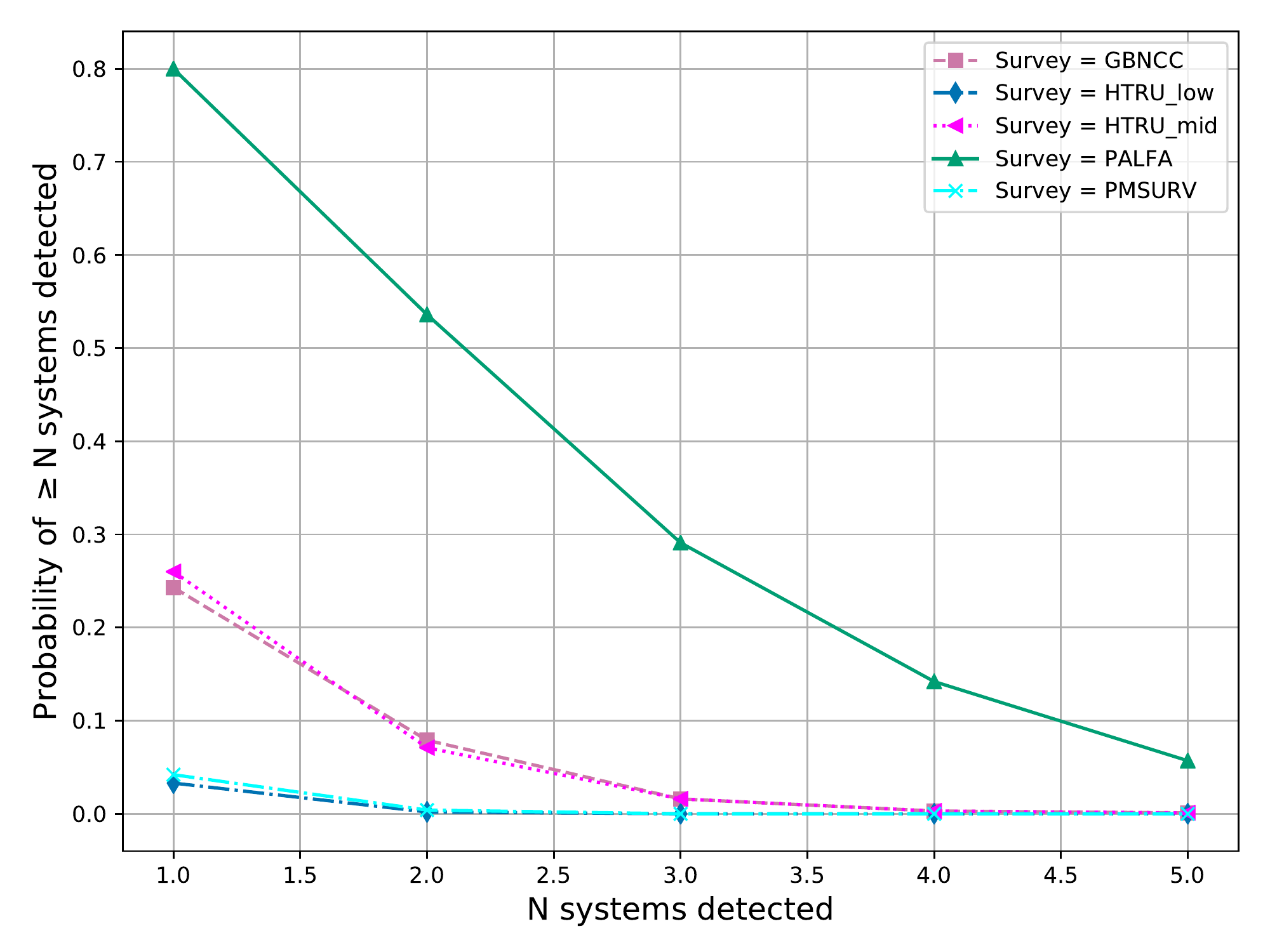}}
        \caption{Complementary cumulative distribution function of the number of UCB systems that are detectable by the radio pulsar surveys listed in Table~\ref{survey_table}, but with integration time set to the optimum value of $\sim$50~s derived in Sec.~\ref{sec_optimum_tint}. The X-axis shows the number of detectable systems, $N$, while the Y-axis shows the probability that $\geq$N systems will be detected in the given survey. Compared to Fig.~\ref{success_survey}, we can see a significant improvement in the detection probability for all the surveys.}
            \label{ideal_success_survey}
    \end{figure*}
    
    Choosing an integration time from the range described above also leads to an average increase in the radiometer S/N, with the biggest effect seen for surveys whose integration times are much higher than the range derived above.  For example, for the PALFA survey, reducing the integration time from 268~s to 50~s increases the S/N of the UCB systems by an average factor of 2.3. On the other hand, for PMSURV, which has the largest integration time of 2100~s, the S/N increases by an average factor of 4.5. The effect of this reduction in the integration time and increase in the S/N on the probability of detecting UCBs is shown in Fig.~\ref{ideal_success_survey}, where we can see a significant increase in the detection probability for the HTRU and PMSURV surveys.
    
    However, given the relatively large range of the optimum survey integration times and the fact that each binary system will have its own optimum integration time, rather than picking a single integration time, we recommend implementing the ``time domain resampling'' technique \citep{jk_time_resampling_srch, ng_accel_search_tech}. In this method, the integration time for a given survey is progressively reduced by a factor of 2 and each chunk of data is searched individually for binary systems.
    Using this method and starting with their design integration times, the survey will be most sensitive to UCB systems when the integration times are between 20~s and 200~s, which correspond approximately to the 95\% confidence limits on the optimum integration time derived above.
    The ``time domain resampling'' method was most recently used in the HTRU survey \citep{ng_accel_search_tech}, but only up to a minimum integration time of $\sim$537~s, which optimized their search to binaries with orbital periods $P_{\rm b} \geq 1.5$~hours. This led to the discovery of J1757--1854, which has an eccentricity, $e = 0.61$ orbital period of $P_{\rm b} = 4.4$~hours and is the most eccentric DNS system detected \citep{1757_accel_search_tech}. 
    Implementing the same time domain resampling technique on all surveys (except AODRIFT, due to its already low integration time) should increase the sensitivity of all the surveys to these UCB systems.
    
\section{Conclusion}
    
    Using the framework developed by \citet{Bagchi_odf}, we develop a neural network to calculate the orbital degradation factor for any given binary system.
    We combine this neural network with \psrpoppy\ opening the possibility for modeling the observed binary pulsar population. We show that, on average, it is easier to detect binary systems which are asymmetric in mass as compared to systems which are symmetric in mass. 
    
    We also investigate the population of UCB systems in the Milky Way as these systems are promising targets for the future space-based gravitational wave observatory LISA. We place upper limits of 1450 and 1100 ultra-compact NS--WD and DNS systems beaming towards the Earth respectively. We also show that the radio pulsar surveys with the Arecibo radio telescope have the highest probability of detecting at least one UCB system. Finally, we show that a survey integration time of $t_{\rm opt} = 42^{+153}_{-22}$~s will maximize the S/N of the UCB systems.
    
    Given the results this work, especially in Sec.~\ref{sec_optimum_tint}, we strongly recommend reprocessing the data collected by the different radio pulsar surveys using the optimum integration time that we derive in this work. The optimum integration times should also be adopted by upcoming radio pulsar surveys at the MeerKAT \citep{meertrap, meertime} and FAST \citep{FAST} radio telescopes. Since the radio pulsar surveys conducted at these telescopes are potentially more sensitive than the Arecibo radio pulsar surveys analyzed in this work, they would have a correspondingly higher probability of detecting an UCB system.
    
\acknowledgements

NP and MAM are members of the NANOGrav Physics Frontiers Center (NSF PHY-1430284). MAM and DRL have additional support from NSF OIA-1458952 and DRL acknowledges support from the Research Corporation for Scientific Advancement and NSF AAG-1616042.
    
\bibliography{bibliography.bib}
\bibliographystyle{aasjournal}

\end{document}